\documentclass[11pt]{article}

\usepackage[divpsnames]{xcolor}
\usepackage{algorithm}
\usepackage{algorithmicx}
\usepackage{algpseudocode}
\usepackage{amsbsy}
\usepackage{amscd}
\usepackage{amsfonts}
\usepackage{amsgen}
\usepackage{listing}
\usepackage{amsmath}
\usepackage{amssymb}
\usepackage{amsthm}
\usepackage{amssymb}
\usepackage{booktabs}
\usepackage{caption}
\usepackage{changepage}
\usepackage[override]{cmtt}
\usepackage{color}
\usepackage{comment}
\usepackage{enumitem}
\usepackage{fancyhdr}
\usepackage{fancyvrb}
\usepackage[T1]{fontenc}
\usepackage[tmargin=1.00in,bmargin=0.75in,lmargin=1in,rmargin=1in,headsep=0.5in]{geometry}
\usepackage{graphicx}
\usepackage{hyperref}
\usepackage{listings}
\usepackage{longtable}
\usepackage{mathpartir}
\usepackage{mathrsfs}
\usepackage{minibox}
\usepackage{multirow}
\usepackage{pgfplots}
\usepackage[section]{placeins}
\usepackage{titlesec}
\usepackage{xspace}
\usepackage{csvsimple}

\usepackage{graphicx}

\theoremstyle{plain}
\newtheorem{theorem}{Theorem}[section]
\newtheorem{lemma}[theorem]{Lemma}

\newtheorem{proposition}[theorem]{Proposition}

\theoremstyle{definition}

\newtheorem{remark}[theorem]{Remark}

\pgfplotsset{compat=1.17} 
\begin{document}

\author{
  Samuel Dittmer\\
  \texttt{samdittmer@stealthsoftwareinc.com}
  \and
  Yuval Ishai\\
  \texttt{yuvali@cs.technion.ac.il}
  \and
  Steve Lu\\
  \texttt{steve@stealthsoftwareinc.com}
  \and
  Rafail Ostrovsky\thanks{Work done while consulting for Stealth}\\
  \texttt{rafail@cs.ucla.edu}
  \and
  Mohamed Elsabagh\\
  \texttt{melsabagh@kryptowire.com}
  \and
  Nikolaos Kiourtis\\
  \texttt{nkiourtis@kryptowire.com}
  \and
  Brian Schulte\\
  \texttt{bschulte@kryptowire.com}
  \and
  Angelos Stavrou\\
  \texttt{astavrou@kryptowire.com}
}
\title{Function Secret Sharing for PSI-CA:\\ With Applications to Private Contact Tracing\footnote{
This research was developed with funding from the Defense Advanced Research Projects Agency (DARPA).  This work was supported by DARPA and NIWC Pacific under contract N66001-15-C-4065 and by DARPA, AFRL/RIKD, USAF, and AFMC under FA8750-18-C-0054. The U.S. Government is authorized to reproduce and distribute reprints for Governmental purposes not withstanding any copyright notation thereon.  The views, opinions and/or findings expressed are those of the author and should not be interpreted as representing the official views or policies of the Department of Defense or the U.S. Government.}}

\date{}

\maketitle

\vfill

\newpage

\section{Introduction}\label{sec:intro}

In this work we describe a token-based solution to Contact Tracing via Distributed Point Functions (DPF)~\cite{EC:GilIsh14} and, more generally, Function Secret Sharing (FSS)~\cite{EC:BoyGilIsh15}.  The key idea behind the solution is that FSS natively supports secure keyword search on raw sets of keywords without a need for processing the keyword sets via a data structure for set membership. Furthermore, the FSS functionality enables adding up numerical payloads associated with multiple matches without additional interaction.  These features make FSS an attractive tool for lightweight privacy-preserving searching on a database of tokens belonging to infected individuals.

More concretely, similarly to the Epione system~\cite{abs-2004-13293}, our proposed solution for decentralized contact tracing securely realizes a variant of the private set intersection (PSI) functionality~\cite{FreedmanNP04}  in the following ``unbalanced'' setting. There are two servers, who each hold a large set of keywords $X$ (infected tokens), and a client who holds a small set of keywords $Y$ (tokens of nearby users). In the basic version of the problem, the client learns the {\em cardinality} of the intersection of $X$ and $Y$ without revealing to any {\em single} server any information about $Y$ (except an upper bound on its size) and without learning any additional information about $X$. (We assume clients to be semi-honest; efficient protection against malicious clients can be obtained using the sketching techniques of~\cite{CCS:BoyGilIsh16}.)
Following~\cite{abs-2004-13293}, we refer to this as {\em PSI cardinality} (PSI-CA). 
We also consider a generalization of PSI-CA in which the client associates to each keyword in $Y$ an integer weight (e.g., a proximity estimate). Here the goal is for the client to obtain the sum of the weights of tokens in the intersection of $X$ and $Y$. We refer to this extended variant as {\em PSI with  weighted cardinality}.\footnote{This is similar to {\em private intersection-sum}~\cite{IonKNPSSSY17}, except that in typical use cases of the latter it is the server who holds the weights. Here both the weights and the output are owned by the client.}

We leverage the capability of FSS-based keyword search to give a lightweight solution for this task. The basic variant of our solution already offers several attractive efficiency features that distinguish it from alternative solutions, including the Epione system~\cite{abs-2004-13293} that uses FSS to realize a similar functionality in a similar setting.
Our solution employs only symmetric cryptography, enabling fast computation and plausible post-quantum security. It involves a single round of interaction consisting of a query from the client to each server followed by a response from each server to the client. The size of the query is comparable to the size of the client's small set $Y$; concretely, in an AES-based implementation the client sends roughly 128 bits for each bit of a keyword in $Y$. The answers are even shorter, and are comparable to the output size. This minimal interaction pattern is particularly useful when the same query is reused for computing intersection with different sets $X$. An incremental variant of our basic solution makes a more fine-grained use of this feature in a setting where both $X$ and $Y$ incrementally change with time.

In terms of computation cost, our basic solution is very fast on the client side: in an AES-based implementation, the client performs roughly 4 AES calls for each bit of each keyword in $Y$. On the server side, the number of AES calls scales linearly with $|X|\cdot |Y|$. While this is good enough for some realistic contact tracing settings, especially when using massive parallelism on the server side (as in the recent FSS-based encrypted search system from~\cite{DORY}), this basic solution does not scale well when the size of $Y$ grows. To improve server computation and make it comparable to $|X|$, one could employ different batching techniques based on hashing or ``batch codes''~\cite{IshaiKOS04,AngelCLS18,CCS:SGRR19,abs-2004-13293}. While these techniques offer a significant improvement in server computation, this comes at the cost of higher communication and setup requirements. 

Instead, we take the following approach. Our starting point is the standard technique of partitioning the keyword domain into buckets, so that {\em on average} only a small number of keywords in $Y$ fall in each bucket. This reduces the PSI task to roughly $|Y|$ instances of secure keyword search, each applying to a single bucket that contains roughly $|X|/|Y|$ elements from $|X|$. Because the FSS outputs are additively secret-shared between the servers, the outputs for different buckets can be summed up without interaction. However, a direct use of this approach requires the client either to reveal the number of keywords in $Y$ that are mapped to each bucket, or alternatively to ``flatten the histogram'' by using dummy queries. The former  results in leaking a small amount of information about $Y$, whereas the latter has a significant toll on performance.  To maximize performance while avoiding leakage, our solution flattens the histogram by deferring keywords from over-populated buckets to be processed with high priority in the next batch of queries.
We use ideas from queueing theory to show that this approach can indeed give superior performance with no leakage, at the price of a very small expected latency in processing queries. 

\subsection{Motivation}

While there are many proposed approaches to contact tracing, most fall short of the privacy and efficiency goals one would desire.  Being a time-sensitive subject, first-to-market solutions have sometimes been marred by privacy concerns.  In this writeup, we explore the usage of FSS to offer attractive new performance and security features compared to other solutions. 

We present a solution that is token-based, decentralized, and customizable with context-sensitive weights (e.g., ``is there a wall between us?'') of tokens, and which prevents clients from directly learning which token infected him or her.  Because it is token-based, we can leverage existing secure decentralized solutions which generate and collect tokens---our key innovation is in the matching rather than collection.  Therefore, using the token-generation API provided by, say, the Apple/Google solution, our matching algorithm would provide strictly greater privacy than simply broadcasting all the infected tokens.

Works such as the Oasis Epione solution~\cite{abs-2004-13293} have considered achieving more security, and~\cite{abs-2004-13293} introduces a new ``private set intersection cardinality'' (PSI-CA) protocol to do so.  The authors give a single-server and a two-server (non-colluding) variant of their solution, and these solutions are two-round protocols secure under the DDH assumption.  Their implementation tradeoffs sacrifice a small (essentially random) amount of privacy such as shard location and hash collisions in order to gain performance.  

In contrast, our solution, which works in the two-server (non-colluding) setting, has the following features:
\begin{itemize}
    \item \textbf{One round.} Our protocol uses only one round, which is surpassed only by solutions that simply broadcast the infected tokens.
    \item \textbf{Minimal cryptographic assumptions.} Our solution relies only on the minimal cryptographic assumption of the existence a secure PRG, which can be instantiated with AES. This gives rise to fast implementations using standard hardware and plausible post-quantum security.
    \item \textbf{Weighted cardinality.} We extend the basic functionality of the PSI-CA primitive into {\em PSI with Weighted Cardinality} that enables a more fine-grained tracing response.
    \item \textbf{Optimal server response size.} Our servers only need to respond to a client query with a single small integer. This is particularly useful in a setting where the same client query is reused for multiple responses.
    \item \textbf{Linear client query size.} The client's queries depend only on the number of tokens the client has seen and does not depend (even logarithmically) on the number of infected tokens seen by the servers.
    \item \textbf{Hashing without leakage via queueing.} Hashing greatly decreases the amount of server work, but it may leak information about the client's queries.  We use queueing theory to delay certain tokens to prevent leakage, but the benefit gained allows us to perform more traces so that a client can check more often overall.    
\end{itemize}

\subsection{Outline}

In Section 2 we give background and related works.  In Section 3 we provide our full design and optimizations. In  Section 4, we introduce techniques to minimize wait times in a streaming solution without leaking, and we provide detailed proofs in Section 5. In Section 6, we provide our full solution and security analysis. In Section 7 we compare our solution to other schemes.  We conclude in Section 8.

\section{Background}\label{sec:background}

\subsection{Related Works on Contact Tracing}

We provide a list of related works in alphabetical order

\begin{itemize}

\item \textbf{Apple and Google}:  \url{https://www.apple.com/covid19/contacttracing/}
\item \textbf{Berkeley Epione}:  \url{https://sunblaze-ucb.github.io/privacy/projects/epione.html}
\item \textbf{Boston University}: \url{https://arxiv.org/abs/2003.13670}
\item \textbf{Catalic}: \cite{AC:DuoPhaTri20}
\item \textbf{Covid-Watch}:  \url{https://covid-watch.org/}
\item \textbf{D3-PT}:  \url{https://github.com/DP-3T/documents}
\item \textbf{Isarel HaMagen}:  \url{https://govextra.gov.il/ministry-of-health/hamagen-app/download-en/}
\item \textbf{Memphis mContain}: \url{https://mcontain.md2k.org/}
\item \textbf{MIT PrivateKit/Safepaths}: \url{https://arxiv.org/abs/2003.08567}
\item \textbf{PEPP-PT}:  \url{https://github.com/pepp-pt/pepp-pt-documentation}
\item \textbf{Singapore TraceTogether}: \url{https://www.tracetogether.gov.sg/}

\end{itemize}

\subsection{Private Set Intersection}

A Private Set Intersection (PSI) protocol~\cite{FreedmanNP04} enables two parties to learn the intersection of their secret input sets $X$ and $Y$, or some partial information about this intersection,  without revealing additional information about the sets. Many variants of this problem have been considered in the literature. We will be interested in {\em unbalanced} PSI, where $|X|\gg|Y|$ and the output should be received by the party holding $Y$, to whom we refer as the {\em client}. We will further restrict the client to learn the {\em size} of the intersection or, more generally, a weighted sum over the intersection, while revealing no other information to the client. 

Most existing PSI protocols from the literature, including protocols based on linearly-homomorphic public-key encryption schemes~\cite{Meadows86,IonKNPSSSY17}, oblivious transfer~\cite{KolesnikovKRT16, PinkasRTY19}, or oblivious linear-function evaluation~\cite{GhoshN19}, are unsuitable for the highly unbalanced case because their communication costs scale linearly with the size of the bigger set $X$. This can be circumvented by PSI protocols that use simple forms of fully homomorphic encryption (FHE)~\cite{ChenLR17,ChenHLR18}. However, FHE-based solutions incur a high computational cost. Moreover, their concrete communication overhead is large when the set $Y$ is relatively small. 

To get around the limitations of traditional PSI techniques, we relax the model by allowing the big set $X$ to be held by two non-colluding servers. In this setting we can get very efficient unbalanced PSI protocols based on the tool of function secret sharing, which we describe next. 

\subsection{Function Secret Sharing}

Our solution heavily builds on the tool of {\em function secret sharing} (FSS)~\cite{EC:BoyGilIsh15}. A (2-party) FSS scheme for a function family $\cal F$ splits a function $f\in\cal F$ into two additive shares, where each share is a function that hides $f$ and is described by a short key. More concretely, a function $f:\{0,1\}^n\to\mathbb{G}$ for some finite Abelian group $\mathbb{G}$ is split into two functions $f_0,f_1$, succinctly described by keys $k_0,k_1$ respectively, such that: (1) each key $k_b$ hides $f$, and (2) for every $x\in\{0,1\}^n$ we have $f(x)=f_0(x)+f_1(x)$. 

We will use FSS for the family $\cal F$ of {\em point functions}, where a point function $f_{\alpha,\beta}$ evaluates to $\beta$ on the special input $\alpha$ and to $0$ on all other inputs. An FSS scheme for point functions is referred to as a {\em distributed point function} (DPF)~\cite{EC:GilIsh14}. We will let ${\sf DPF.Gen}(1^\lambda,\alpha,\beta)$ denote the DPF key generation algorithm, which given security parameter $\lambda$ and the description of a point function $f_{\alpha,\beta}$ outputs a pair of keys $(k_0, k_1)$ (where here we assume for simplicity that the group $\mathbb G$ is fixed). 
We use ${\sf DPF.Eval}$ to denote the evaluation algorithm that on input $(k_b,x)$ returns an output share $y_b$ such that $y_0+y_1=f_{\alpha,\beta}(x)$. 

We rely on the best known DPF construction from~\cite{CCS:BoyGilIsh16}, which has the following performance features with an AES-based implementation: The length of each key is roughly $128n$ bits (some savings are possible when the group $\mathbb G$ is small). The cost of ${\sf DPF.Gen}$ is roughly $4n$ AES calls, whereas the cost of ${\sf DPF.Eval}$ is roughly $n$ AES calls, where both can be implemented using fixed-key AES.

A direct application of DPF for secure keyword search in a 2-server setting was suggested in~\cite{EC:GilIsh14,EC:BoyGilIsh15}. Secure keyword search can be viewed as an extreme instance of unbalanced PSI where $|Y|=1$. Here we generalize this in two dimensions: first, we allow a client to have multiple keywords, thus supporting a standard PSI functionality. We propose different methods for improving the cost of independently repeating the basic keyword search solution for each keyword in the client set $Y$. Second, we exploit the ability to use a general group $\mathbb G$ for implementing a {\em weighted} variant of PSI where each of the client's secret keywords has an associated secret weight. In fact, we use a product group for revealing multiple weighted sums.

\section{Design}\label{sec:design}

Following the Epione system of Trieu et al.~\cite{abs-2004-13293}, we capture a private contact tracing functionality as a variant of PSI Cardinality, namely privately computing the size of the intersection between a set of tokens collected by a client's phone and a set of tokens belonging to infected patients. We consider here the 2-server setting, whose overhead is smaller by orders of magnitude than similar 1-server solutions.  

We extend the PSI Cardinality functionality from~\cite{abs-2004-13293} in two ways. First, we allow the client's tokens to have weights. These weights are represented as an abstract group element $G$ which can, for example, be a product group that packs various slots of factors depending on available sensors, etc.  We abstract this out to have the client compute a single ``risk score'' represented as a scalar. We refer to this extended functionality as ``PSI with Weighted Cardinality,'' or PSI-WCA for short. Second, we we use FSS directly to allow for a \emph{one-round---one up and one back---}solution that supports both the ``one-shot'' version, with a single pair of input sets, and an ``incremental'' version, which takes advantage of the fact that only a small fraction of the inputs on each side changes in each time period. 

The following subsections are organized as follows. We start by formalizing the functionality and presenting a basic solution for the one-shot case. We then describe an improved solution for the incremental case. Finally, we discuss several kinds of optimizations that can apply to both the one-shot and the incremental case. 

\subsection{The One-Shot Case}
\label{ss:oneshot}

The functionality we realize is an extended ``weighted'' version of PSI Cardinality that attaches a weight to each client item. 
\medskip

\noindent 
{\bf Functionality PSI-WCA}:
\begin{itemize}
\item {\sc Inputs:} 
\begin{itemize}
\item Each of the two servers $S_0,S_1$ holds the same set $X=\{x_1,\ldots,x_N\}$ of $k$-bit strings referred to as {\em tokens}.
\item Client holds a set $Y$ of pairs of the form $Y=\{(y_1,w_1),\ldots,(y_n,w_n) \}$, where each $y_i$ is a $k$-bit token and each $w_i$ is an element of an Abelian group $G$ (typically we choose to work over the integers with large enough modulus to prevent wraparound, but using an arbitrary group allows for the ability even to support product groups with multiple slots encoding different pieces of information).
\end{itemize}
\item {\sc Outputs:} Client outputs the sum of the weights of the tokens in the intersection; namely, the output is $w=\sum_{i:y_i\in X} w_i$ where summation is in the group $G$. We can handle maliciously formed inputs using verifiable FSS ideas, though this still does not prevent a client from picking arbitrary inputs.  In order to address these simultaneously, we can rely on a Trusted Execution Environment (TEE) on a client's device to store tokens and perform these operations.
Servers have no output. 
\item {\sc Leakage:} The size parameters leaked to the adversary are $k,n,G$.
\end{itemize}

\paragraph{Typical parameters.} Trieu et al.~\cite{abs-2004-13293} suggest that the number of tokens collected {\em daily} on the server side is $N=6\cdot 10^6$ and on the client side is $n=80$. Both should be multiplied by 14 when aggregating over a 2-week period. The raw token length is $k=128$, but it can be pruned to $k=74$ or bits without incurring a significant error probability. For the weighted case, we let $G=\mathbb Z_{2^{16}}$ to accommodate integer weights with output size bounded by $2^{16}$.

\paragraph{The baseline solution.} We follow the approach of Boyle et al. for secure keyword search via a direct use of distributed point functions (DPFs)~\cite{EC:GilIsh14,CCS:BoyGilIsh16}. This departs from the approach of Chor et al.~\cite{ChorGN98} and Trieu et al.~\cite{abs-2004-13293}, which uses a data structure (Cuckoo Hashing in~\cite{abs-2004-13293}) for reducing keyword search to private information retrieval (PIR). The direct DPF-based approach requires one round of interaction and accommodates the weighted case with almost no extra overhead.  

While we describe the protocol using direct interaction of the client with the two servers $S_0,S_1$, in practice it may be preferable to have the client interact only with $S_0$ and have (encrypted) communication to and from $S_1$ routed via $S_0$. In the following we use $\lambda$ to denote a security parameter, and we consider security against a {\em passive} (aka semi-honest) adversary corrupting either one of the two servers or the client. 

\medskip

\noindent 
{\bf Protocol PSI-WCA}:
\begin{itemize}
\item {\sc Client-to-servers communication:} 
\begin{enumerate}
\item
For each client input pair $(y_i,w_i)$, Client generates a pair of DPF keys $(k^0_i, k^1_i)\leftarrow {\sf DPF.Gen}(1^\lambda,y_i,w_i)$.
\item Client sends the $n$ keys $k^b_i$ to server $S_b$. 
\end{enumerate}
\item {\sc Servers-to-client communication:} 
\begin{enumerate}
\item Each server $S_b$ computes $a'_b:=\sum_{j=1}^N \sum_{i=1}^n {\sf DPF.Eval}(k^b_i,x_j)$, where summation is in $G$. (Each such invocation of ${\sf DPF.Eval}$ can be implemented with roughly $k$ invocations of fixed-key AES and does not require any communication between servers.)
\item Letting $r\in_R G$ be a fresh secret random group element shared by the two servers, $S_0$ sends to Client $a_0:=a'_0+r$ and $S_1$ sends $a_1:=a'_1-r$, where addition and subtraction are in $G$.  This can be generated using a shared pseudorandom sequence known only to the servers (e.g., a common PRF seed).
\end{enumerate}
\item {\sc Client output:} Client outputs $w=a_0+a_1$, where summation is in $G$.
\end{itemize}

The correctness of the above protocol is easy to verify. Security against a single server follows directly from the security of the DPF. Security against the Client follows from the blinding by $r$, which makes the pair of answers received by Client random subject to the restriction that they add to the output. 
We now discuss the protocol's efficiency.

\paragraph{Performance.} Using an AES-based implementation of the DPF from~\cite{CCS:BoyGilIsh16}, the above protocol has the following performance characteristics:
\begin{itemize}
\item {\sc Rounds}: The protocol requires a single round of interaction, where Client sends a query to each server $S_b$ and gets an answer in return. Client's query can be reused when the client's input $Y$ does not change, even when the server input $X$ changes.
\item {\sc Communication}: Client sends each server $\approx 128\cdot kn$ bits and gets back a single element of $G$ from each server.
\item {\sc Computation}: Client performs $\approx 2kn$ (fixed-key) AES calls to generate the queries. The cost of reconstructing the answer is negligible.  The computation performed by each server is dominated by $\approx knN$ AES calls. For modern processors (see Footnote 12 of~\cite{abs-2004-13293}), each AES call requires 10 machine cycles, which enables $360\cdot 10^6$  AES calls per second on a 3.6 GHz machine. This can be further sped up via parallelization.
\end{itemize}

\subsection{Incremental Mode}

The incremental mode captures a dynamic ``streaming'' version of the problem where the sets $X$ and $Y$ held by the servers and the client change in each time epoch (say, each day) by $N'$ and $n'$ respectively. We typically consider $N'\ll N$ and $n' \ll n$. There is a time period of $T$ epochs (say, $T=14$) by which tokens expire.  We describe a better streaming design at the end of this section. The goal is to compute the PSI-WCA functionality in the sliding window corresponding to each epoch, where the inputs consist of the $N=TN'$ and $n=Tn'$ tokens collected during the last $T$ epochs by the servers and client, respectively.  

In this incremental mode, we let the client generate and communicate new queries only for the $n'$ tokens introduced in each epoch. These queries are stored on the server side for $T$ epochs, and are erased once they expire. In each epoch, the servers only need to match the new $n'$ client tokens with all $TN'$ server tokens and the new $N'$ server tokens with all $n'T$ client tokens. The incremental mode reduces the number of AES calls per epoch on the client side from $Tkn'$ to $kn'$, and on the server side from $knN=T^2kn'N'$ to roughly $kT\cdot (n'N+nN')$.  The client communication and computation per epoch are each reduced by a factor of $T$ compared to the one-shot solution.

\subsection{Optimizations}

We now describe different optimizations and efficiency tradeoffs that allow one to reduce costs on the server and/or client side, typically at the expense of a milder increase in other costs and a small amount of leakage to the client beyond the output of PSI-WCA. 

\paragraph{Improving server computation via hashing.} Similarly to the simple use of hash functions and batch codes for amortizing the server computation of multi-query PIR~\cite{IshaiKOS04,AngelCLS18}, and similar techniques for standard PSI, one can use a similar approach for amortizing the server computation in PSI-WCA. The idea is to randomly partition the token domain into a small number of buckets $\ell$ via a public hash function $H:\{0,1\}^k\to[\ell]$ (typically $\ell \approx n$), and let the client match each token $y_i$ only with the tokens in bucket $H(y_i)$. To make this possible, we need the client either to reveal the number of tokens $y_i$ mapped to each bucket (which leaks a small amount of information about $Y$ to the servers) or to add dummy tokens $y^*_j$ to ensure all buckets have a fixed size except with small failure probability. Compared to more sophisticated data structures such as Cuckoo hashing, discussed next, this approach does not require additional interaction and is suitable to the incremental mode in which new server tokens are added on the fly.   
 
\paragraph{Improving server computation via data structures.} Trieu et al.~\cite{abs-2004-13293}, following a more general approach of Chor et al.~\cite{ChorGN98}, employ a Cuckoo hashing data structure to reduce the keyword search problem (of matching a single client token $y_i$ with all $N$ tokens $x_j$) to two invocations of PIR on a $2N$-bit database. The main advantage of this approach over our baseline solution is that, using the efficient DPF $\sf EvalAll$ procedure from~\cite{CCS:BoyGilIsh16}, the number of AES invocations on the server side is reduced by roughly a factor of $k/4$. However, compared to our more direct approach, this makes the solution much more complex. 
In particular, it requires an additional round of interaction and a bigger answer size and, perhaps most significantly, is not compatible with our incremental mode. For data sizes in which this approach is attractive despite the above disadvantages, we propose two additional optimizations that were not considered in~\cite{abs-2004-13293}. The first is to apply an ``early termination'' procedure suggested in~\cite{CCS:BoyGilIsh16} to further reduce the number of AES calls on the server side by an additional factor of 64. The second is to amortize the cost of multiple PIR instances via (deterministic or probabilistic) batch codes~\cite{IshaiKOS04,AngelCLS18}, which can additionally reduce the server computation by up to a factor of $n/2$. 

\paragraph{Trading token length for answer size.}  A third type of optimization, which can reduce the work of both servers and clients by roughly a factor of 2, is to reduce the token length in a way that may give rise to false positives, but to provide at the same time a mechanism for detecting such false positives. This optimization can be applied on top of the baseline solution or its hashing-based optimization, without incurring the disadvantages of the data structures approach. The starting point is the observation that the token length $k$ appears as a multiplicative term in all complexity measures. While the concrete size of $k$ is not too big (Trieu et al.~\cite{abs-2004-13293} suggest pruning 128-bit tokens to $k=74$ bits), further reducing token size, say to $k'=40$, can directly improve all cost measures. A straightforward approach is simply to hash $k$-bit tokens to a smaller size $k'$; in fact, assuming tokens are pseudorandom, this can be done via simple truncation. Let $X',Y'$ denote the sets of truncated tokens. The problem with making $k'$ too small is the $2^{-k'}$ probability of a false positive for each attempt to match a client token with a server token. The probability of false positives can be reduced by providing a cheap mechanism for detecting the existence of false positives. If we make the assumption that a non-empty intersection is small, we can aggregate the information about {\em full $k$-bit tokens} corresponding to the intersection of (truncated) sets $X'$ and $Y'$ by using standard linear sketching techniques, while incurring a small {\em additive} overhead of $O(k)$ to the query and answer size, and with only a small additive computational overhead. 
Concretely, the client generates its query using the set $Y'$ of $k'$-bit tokens, but with a bigger DPF group $G'=G\times H$, where $H$ is the output domain of a suitable linear sketching function for set membership. The servers append to the $G$-component of their answer, computed using the $k'$-bit token set $X'$, an $H$-component obtained by mapping each $k$-bit token $x_i$ to an element $h_i$ from $H$. The sketching has the property that the client can distinguish between a sketch aggregating a bounded (nonzero) number of tokens from $Y$ from one that corresponds to a false positive. Examples for suitable sketches include Bloom filters, power-sum sketches~\cite{CorriganGibbsB15}, or the probabilistic sketches from~\cite{OstrovskyS07}. Optimizing the efficiency of this approach while minimizing the amount of additional leakage remains to be further explored. 

\subsubsection{Streaming and bucketing}
\label{ss:bucketing_design}

We give a rough overview of the various approaches here, with a more mathematical analysis of the expected wait times in Section~\ref{sec:wait}.

We have $N$ infected (70k * 60/5 tokens per hour * 24 hours a day) with $n$ on a users phone (50k tokens over 2 weeks). The naive secure method  requires 
\begin{itemize}
    \item Client Work: $n$ FSS gens
    \item Communication: $n$ FSS keys
    \item Server Work: $n*N$.
\end{itemize} The straightforward insecure solution, a linear scan, requires
\begin{itemize}
    \item Client Work: 0
  \item Communication: $n$ tokens
  \item Server Work: $n+N$ 
\end{itemize}
  
To get something closer to the insecure solution in cost, we use a bucketing solution. We start with $m$ buckets of some bin size $b$, hash all $n$ things into the $m$ buckets. If we choose $b$ large enough, then except with the some failure probability $\varepsilon < 2^{-40}$ (say), all tokens are assigned to a bucket. The server then hashes each of their $N$ tokens and checks against the $b$ values in the corresponding bucket. This reduces server work to $N * b$, while increasing communication to $m * b$, but we can choose a value of $m$ such that~$m * b$ is close to $n$. For additional efficiency improvements, we can make $b$ smaller and allow buckets to overflow, moving all overflow tokens into a stash that carries over to the next day. 

We have choices here to make about how we hash: we can use the same hash function each day, or we can refresh the hash function each day, and we can use a single hash function or multiple hash functions. When we use $c > 1$ hash functions, we use a greedy algorithm to assign each token to whichever bucket is currently the most empty. This increases server work to $N * b * c$, but allows the stash to be much smaller, and so reduces expected wait time.

We have performed Monte Carlo simulations of this procedure to get estimates of the expected wait time, which we compare to the theoretical steady state expected wait wait time in Table~\ref{tab:experiments}.

 \begin{remark} If you do leaky balancing: server picks a hash function that makes things as equal as possible, another leaky possibility is Epione solution where the first few bits is the hash.
\end{remark}

\section{Minimizing wait times in a streaming solution}
\label{sec:wait}

\subsection{Setting}

One key drawback of a streaming solution is that some tokens will take longer than one day to be processed. Additionally, as time passes, the backlog of unprocessed tokens builds up, and the  wait time increases. To understand the tradeoffs involved, we analyze the expected average and worst case wait times. When we choose parameters appropriately, the backlog in the stash reaches a steady state of reasonable size, the average wait time is small, and very large wait times are extraordinarily rare.

In our analysis, we consider two metrics under four scenarios.  We measure \textit{expected wait time} and \textit{expected worst-case wait time}, both once a steady state has been reached. Formally, the first metric is the limit as $t \to \infty$ of the expectation of the average wait time over all tokens inserted at time $t$, while the second is the limit as $t \to \infty$ of the expectation of the maximum wait time over all tokens inserted at time~$t$. We consider the first metric in the limit as $n \to \infty$, while the second we consider as a function of $n$, since the probability of extraordinarily rare events increases with the sample size.

The four scenarios we consider are (i) Fixing $c=1$ hash function to distribute tokens, (ii) Refreshing the $c=1$ hash function each day, (iii) fixing $c > 1$ hash functions and (iv) refreshing $c > 1$ hash functions each day. 

For each scenario, we consider parameters $n$, the number of tokens, $m$ the number of buckets, $b$ the bin size, and the occupancy ratio $\alpha := n/(bm)$. Additionally we have $c$, the number of hash functions, and $R$, a single bit representing whether or not we re-randomize each day. Theoretical results as $n \to \infty$ depend on $(b,c,\alpha, R)$, while for experimental results we give additionally the parameter $n$. We compare our experiments with the steady state wait time as $n \to \infty$ in Table~\ref{tab:experiments} and give an overview of asymptotic results in Table~\ref{tab:asymptotics}.

\subsection{Results}

\subsubsection{Summary of results}

The bounds on expected wait times and expected worst-case wait times we give here are primarily calculations using existing work. Proposition \ref{prop:cuckoodifferent} is an extension of work by Mitzenmacher \cite{mitzenmacher1999studying}.

\begin{itemize}
    \item The expected wait time decreases exponentially with $b$ for $c=1$ hash function, and doubly exponential with $b$ for $c > 1$ hash functions.
    \item The expected worst-case wait time is $\Theta(\log n)$ for each scenario except $c > 1$ fixed hash functions, where it is $\Theta(\log \log n)$.  
    \item As $\alpha$ increases, the expected wait time for the re-randomizing solution decreases relative to the fixed hash function solution. We give in Table~\ref{tab:alpha_eq} the value of $\alpha$ where the two solutions match exactly in expected wait time, for various choices of $b$ and $c$. 
\end{itemize}

\subsubsection{Rerandomization of hash function, \texorpdfstring{$c=1$}{c=1}}

\begin{proposition}
	\label{prop:onehashdifferent}
	When~$c=1$ and the rerandomization bit $R=\textsc{True}$, and $e\alpha < 1$, $$\mathbb{E}[W] \leq (e\alpha)^{-b}$$ and $$\mathbb{E}[\max{W}] = - \frac{\log n}{b\log \alpha} + O(1).$$
\end{proposition}

\subsubsection{Fixed hash function, \texorpdfstring{$c = 1 $}{c = 1}}

\begin{proposition}
	\label{prop:onehashsame}
	When~$c=1$ and the rerandomization bit $R=\textsc{False}$, $$\mathbb{E}[W] \leq \frac{\alpha^{b} e^{(1-\alpha)b}}{1 - \alpha^{b} e^{(1-\alpha)b}}$$ and $$\mathbb{E}[\max{W}] = - \frac{\log n}{b\log \alpha} + O(1).$$
\end{proposition}

\subsubsection{Rerandomization of hash function, \texorpdfstring{$c > 1$}{c > 1}}

\begin{proposition}
	\label{prop:cuckoodifferent}
	When~$c > 1$ and the rerandomization bit $R =\textsc{True}$, and $\alpha$ and $b$ are chosen such that~$0 < \alpha b < 1$, then $$\mathbb{E}[W] \leq \left(\alpha b\right)^{c^b}$$ and $$\mathbb{E}[\max{W}] = \frac{\log n}{-c^b \log (\alpha b)} + O(1).$$
\end{proposition}

\begin{remark}
    This result may hold without any restriction on $\alpha$ besides $\alpha < 1$, for $b$ sufficiently large (where the lower bound on~$b$ is a function of~$\alpha$). But via a simple heuristic, for $\alpha = 1-\varepsilon$ we require at least $b > \frac{1}{\varepsilon}$ before the doubly exponential bounds could be effective. In practical situations, keeping $\alpha$ and $b$ small will lead to more efficient implementations.
\end{remark}

\subsubsection{Fixed hash function, \texorpdfstring{$c  > 1 $}{c > 1}}

\begin{proposition}
	\label{prop:cuckoosame}
	When~$c > 1$ and the rerandomization bit $R =\textsc{False}$, and $\alpha$ and $b$ are chosen such that~$0 < \alpha b < 1$, then $$\mathbb{E}[W] = O\left((\alpha b)^{c^b-1}\right)$$ and $$\mathbb{E}[\max{W}] = \frac{\log \log n}{\log c} + O(1)$$
\end{proposition}

\begin{table}[htbp]
\begin{tabular}{|l|l|l|l|l|}
\hline
Wait time for $(\alpha, b, n)$          &  $(0.313,2, 25000)$ & $\lim_{N \to \infty}(0.313,2, n)$ & $(0.417,3,25000)$ &  $\lim_{n \to \infty} (0.417,3,n)$ \\ \hline
$c = 1, R = \textsc{True}$ &    0.05319 & 0.05567                       &    0.04512                   &  0.04604         \\ \hline
$c=1, R=\textsc{False}$ & $0.05904$         & $0.06022$  &                   0.04961  &  0.05063                       \\ \hline
$c = 2, R = \textsc{True}$ &  0.00073  &     0.00075                  &  0.00009                       & $0.00008$       \\ \hline
$c = 2, R=\textsc{False}$ & 0.00076        &    0.00074                     & 0.00007                       &  0.00008       \\ \hline
\end{tabular}
\caption{Experimental and theoretical wait times}
\label{tab:experiments}
\end{table}

\begin{table}[htbp]
    \begin{minipage}{.5\linewidth}
      \centering
        \begin{tabular}{|l|l|l|}
\hline
Wait time & Average & Worst-case \\ \hline
$c = 1, R = \textsc{True}$ &    $(\alpha e)^{-b}$       &    $\frac{\log n}{-b \log \alpha} + O(1)$        \\ \hline
$c=1, R=\textsc{False}$ & $\frac{\alpha^{b} e^{(1-\alpha)b}}{1 - \alpha^{b} e^{(1-\alpha)b}}$         &          $O(\log n)$  \\ \hline
$c > 1, R = \textsc{True}$ &   $\left(\alpha b \right)^{c^{b}}$      &   $\frac{\log n}{-c^b \log(\alpha b)} +O(1) $       \\ \hline
$c > 1, R=\textsc{False}$ &     $O\left(\left(\alpha b \right)^{c^{b}-1} \right)$        & $\frac{\log \log n}{b \log c} + O(1)$           \\ \hline
\end{tabular}
\caption{Asymptotic wait times, as a function of $(\alpha, b, c)$}
\label{tab:asymptotics}
    \end{minipage}%
    \quad \quad \quad \quad \quad
    \begin{minipage}{.32\linewidth}
      \centering
        \begin{tabular}{|l|l|l|}
\hline
$\alpha_{\textsc{eq}}(b)$ & $c=1$ \\ \hline
$b=1$ & 0.63890      \\ \hline
    $b=2$ & 0.43318  \\ \hline
    $b=3$ & 0.31706  \\ \hline
$b=4$ & 0.24632         \\ \hline
\end{tabular}
    \caption{Value of $\alpha$ where fixed hash matches rerandomization}
    \label{tab:alpha_eq}
    \end{minipage} 
\end{table}

\subsection{Probabilistic models} 

A single round of the token distribution procedure can be analyzed as a classic balls-and-bins problem, with $m$ bins and $n = b\alpha m $ balls to distribute at random among those bins. In the token distribution procedure, we have bins of size $b$, remove all balls from all bins at the end of each day, and cache the overflow elements. For the purposes of analyzing the expected wait time and expected cache size, it is equivalent to allow the bins to have infinite size and remove $b$ balls from each bin. The expected number of balls remaining in the bins in this setting is equal to expected cache size in the token distribution procedure.

When we re-randomize the hash functions each day, then, in the limit as $n \to \infty$, the daily ratio of the stash size to $n$ forms a Markov chain on a continuous state space. We can compute the steady state from the transition probabilities, and the expected wait time by Little's law.

With $c=1$ fixed hash function, as $n \to \infty$, the  distributions of new balls placed in each bin are independent and follow a Poisson distribution with parameter $\alpha b$. Each bin's behavior therefore matches a discrete time GI-D-c queue (with the $c$ in the queue definition equal to our $b$ defined here) since processing $b$ balls at once is equivalent to having $b$ servers with a fixed processing time. The steady state distribution of this queue was first derived in \cite{bruneel1994analysis}, see also \cite{janssen2005analytic} for a survey of prior work and additional analytical tools.

When there are $d > 1$ fixed hash functions, the distributions of distinct bins are no longer independent. The remarkable $O(\log \log n)$ bound on worst-case wait time first appeared in \cite{azar1994balanced} as a bound on bin size. We study the wait times using the differential equation method of Mitzenmacher \cite{mitzenmacher1999studying}.

\section{Wait time proofs}

\subsection{Rerandomization}
We mention briefly some general techniques that apply to the rerandomization regime, both in the $c=1$ and the $c > 1$ case.

Performing the distribution procedure on~$\alpha bn$ balls together with~$(\beta-\alpha) b n$ balls leftover from the previous round is equivalent to performing the distribution on $\beta b n$ balls.

The steady state solution therefore occurs at the value of~$\beta$ for which~$\alpha b n$ balls are removed after applying the distribution procedure on $\beta b n$ balls. At the steady state, the probability that an individual ball is removed is equal to
$
\frac{\alpha}{\beta}.
$
Each round of the steady state is independent, since we choose a new hash function each time, so the amount of time a ball spends before being removed is distributed geometrically, with mean
$
\mathbb{E}[W] = \frac{\beta}{\alpha} - 1.
$

Via a standard calculation (see e.g. \cite{eisenberg2008expectation}), when there are $n$ bins we have $$\mathbb{E}[\max{W}] = \frac{\log n}{\log\left(\frac{\beta-\alpha}{\alpha}\right)} + O(1).$$

\subsection{\texorpdfstring{$c=1$}{c = 1} hash function}

\subsubsection{One hash function: Proof of Proposition~\ref{prop:onehashdifferent}}

Writing~$a_{t,k}(\beta)$ to indicate the dependence of the distribution on~$\beta$, we have
$$
\alpha b = \sum_{k=1}^{b} k a_{t,k}(\beta) + b \sum_{k > b} a_{t,k}(\beta).
$$
Since the~$a_{t,k}$'s sum to~$1$, we obtain:
$$
\alpha = 1 - \frac{1}{b} \sum_{k=0}^{b} (b-k) a_{t,k}(\beta).
$$

This is a Poisson process, so we have
$$
a_{t,k} = e^{-b\beta} \frac{(b\beta)^k}{k!}
$$
and
\begin{align*}
\alpha &= 1 - \frac{e^{-b\beta}}{b} \sum_{k = 0}^{b} \frac{(b-k)b^k\beta^k}{k!} \\
&= 1 - e^{-b\beta} \sum_{k = 0}^{b} \frac{b^k\beta^k}{k!} + \beta e^{-b\beta} \sum_{k = 0}^{b} \frac{b^{k-1}\beta^{k-1}}{(k-1)!} \\
&= 1 -e^{-b\beta} \frac{b^{b}\beta^{b}}{b!} - e^{-b\beta}(1-\beta) \sum_{k = 0}^{b-1} \frac{b^k\beta^k}{k!} 
\end{align*}

For~$b > 1$, by Taylor's Approximation, we have
$$
1-e^{-b\beta}\frac{(b\beta)^b}{b!} - e^{-b\beta}(1-\beta)(e^{b\beta} - \frac{(b\beta)^b}{b!}) \leq \alpha \leq 1-e^{-b\beta}\frac{(b\beta)^b}{b!} - e^{-b\beta}(1-\beta)(e^{b\beta} - e^{b\beta}\frac{(b\beta)^b}{b!})
$$
$$
\beta\left(1 - \frac{e^{-b\beta}(b\beta)^b}{b!}\right)\leq \alpha \leq \beta +(1-\beta-e^{-b\beta})\frac{(b\beta)^b}{b!}
$$

By Stirling's Approximation, for~$\beta < \frac{1}{e}$, the expression~$(b\beta)^b/b!$ is exponentially small in~$b$, so that~$|\alpha-\beta|$ is bounded above by a quantity exponentially small in~$b$.

\subsubsection{One hash function: Proof of Proposition~\ref{prop:onehashsame}}

This is a GI-D-c queue, with $b$ servers per queue, and the inputs following a Poisson distribution. 

Combining \cite{janssen2005analytic} with Little's law \cite{little1961proof}, we can write the expected wait time as $$\mathbb{E}[W] = \frac{1}{\alpha b}\sum_{\ell \geq 1} \frac{1}{\ell}\sum_{i > \ell b}{(i - \ell b) \frac{e^{-\ell \alpha b}(\ell \alpha b)^i}{(i)!}}.$$ The sum of $\frac{e^{-\ell \alpha b}(\ell \alpha b)^i}{i!}$ for $i > \ell b$ is equal to $R_{\ell b}[A^{\ell}(z)]$, that is, to the $\ell b$th remainder of the Taylor polynomial for $(A(z))^{\ell}$, where $A(z) = e^{\alpha b(z-1)}$. By the integral form of the remainder, this is equal to $$\int_{0}^1  \frac{(\ell \alpha b)^{\ell b}}{(\ell b)!}(1-t)^{\ell b} e^{\ell \alpha b(t-1)} dt.$$

Similarly, the sum of $i \frac{e^{-\ell \alpha b}(\ell \alpha b)^i}{i!}$ is equal to $R_{\ell b - 1}[\frac{d}{dz}(A^{\ell}(z))]$, which is equal to
$$\ell \alpha b \int_{0}^1  \frac{(\ell \alpha b)^{\ell b-1}}{(\ell b-1)!}(1-t)^{\ell b-1} e^{\ell \alpha b(t-1)} dt.$$

Combining, we have
\begin{align*}
\mathbb{E}[W] &=  \frac{1}{\alpha b}\sum_{\ell \geq 1} \int_{0}^1  \frac{(\ell \alpha b)^{\ell b-1}}{(\ell b-1)!}(1-t)^{\ell b-1} e^{\ell \alpha b(t-1)} \left(\alpha b - \frac{b \ell \alpha b}{\ell b}(1-t)\right) dt \\
&=   \sum_{\ell \geq 1} \frac{(\ell \alpha b)^{\ell b}}{(\ell b - 1)!} \int_{0}^{1} t(1-t)^{\ell b - 1} e^{\ell \alpha b(t-1)} dt
\end{align*}
As $b \to \infty$, by Stirling's formula the log of the term outside the integral goes to $\ell b (1 + \log \alpha)$. The log of the integrand is $\ell b (\alpha(t-1)) + \ell b(\log(1-t))$. Combining these two terms gives $$\ell b\left(-u+(1-\tfrac{1}{\ell b})\log u + 1\right),$$, for $u = \alpha(1-t)$. This expression is monotone, and we can bound the~$(1-\tfrac{1}{\ell b})$ term below by~$(1 - \tfrac{1}{b})$, so we have $$(1+(1 + \tfrac{1}{b} \log u - u) < (1+\log \alpha - \alpha).$$ Summing the resulting geometric series over~$\ell$ gives $$\mathbb{E}[W] \leq \frac{\alpha^{b} e^{(1-\alpha)b}}{1 - \alpha^{b} e^{(1-\alpha)b}}$$ as desired. The proof of the bound on~$\mathbb{E}[\max W]$ follows by a similar anaylsis of the expression for the pgf of the stationary distribution given in~\cite{janssen2005analytic}.

\subsection{\texorpdfstring{$d > 1$}{d > 1} hash functions}

Using the approach in Mitzenmacher, we define $c_k(t)$ to be the proportion of bins carrying~$k$ balls after having distributed~$tn$ of the balls, and define
$$
s_i(t) := \sum_{j \geq i} c_j(t).
$$
Then the~$s_i$'s satisfy the differential equations
$$
\frac{d s_i}{dt} = s_{i-1}^2-s_i^2,
$$
and we note that~$s_0$ is identically equal to one. 

\subsubsection{\texorpdfstring{$d > 1$}{d > 1} hash functions, fixed} We remove~$b$ balls at time~$b\alpha$, giving us the steady-state equation
$$
s_i(b\alpha) = s_{i-b}(0),
$$
for~$i > b$. We also have
$$
\sum_{i=1}^{b-1}s_i(b\alpha) = b\alpha,
$$
because in the steady state~$b\alpha n$ balls are added and removed each round. Adapting the method of Mitzenmacher \cite{mitzenmacher1999studying}, we get double exponential bounds on stash size, which gives the desired bounds on expected wait time and expected worst-case wait time.

We have $$b s_i(0) < \sum_{j=i-b+1}^{i} s_j(0) \leq b\alpha (s_i(b\alpha))^d = b\alpha (s_{i-b}(0))^d$$ and $$s_i(0) - s_{i+b}(0) \leq b\alpha (s_{i-1}(0))^d,$$ from the integral equations. Combining these gives $s_i(0) < \frac{b \alpha}{1-\alpha^d} (s_{i-1}(0))^d$, and by induction we have $s_i(0) < b\alpha \left(\frac{b \alpha}{1-\alpha^d} (s_{1}(0))^d\right)^{d^{i-2}}$. Similarly we show $s_i(t) \leq (s_1(t))^{d^{i-1}} \leq (b\alpha)^{d^{i-1}}$. The steady state equation is $$b \alpha = \sum_{i=1}^b s_i(0) + \int_0^{b\alpha} 1 - s_b^d(t) dt.$$ Applying the above bounds gives $s_1(0) < b\alpha (b\alpha)^{d^b}$. By the doubly exponential decay of $s_i(0)$ with respect to $i$, the steady state stash size is equal to $(1+o(1))b\alpha (b\alpha)^{d^b}$, and the desired bound on $\mathbb{E}[W]$ follows from Little's law. The long term probability any party has a wait time of at least $k$ is equal to $$\sum_{i> bk} s_{i}(0)$$, so the probability that the maximum wait time is at least $k$ is $$1 - \left(1 - (b\alpha)^{d^{bk}}\right)^n$$. For $k = \frac{\log \log n}{b \log d} + O(1)$, this is exponentially small in $n$, which completes the proof for $\mathbb{E}[\max W]$.

\subsubsection{\texorpdfstring{$d > 1$}{d > 1} hash functions, rerandomization: Proof of Proposition~\ref{prop:cuckoodifferent}}

When~$b=1$, the number of balls removed is~$$n \sum_{i=1}^b s_i(\beta),$$ so, as in the proof of Proposition~\ref{prop:onehashdifferent} we seek to determine the value of~$\beta$ such that 
$$
\alpha b = \sum_{i=1}^b s_i(\beta).
$$
From the differential equations and induction, it follows that
$$
s_i(t) \leq (\beta b)^{(d^i-1)/(d-1)}.
$$
From the definition of $s_i$, we have $$ \frac{d}{dt} \sum_{i > b} s_i(t) = (s_b(t))^d  \leq (\beta b)^{(d^b-1)\left(1 + \tfrac{1}{d-1}\right)} \approx (\alpha b)^{(d^b-1)\left(1 + \tfrac{1}{d-1}\right)}.$$

Since $\beta b$ balls total are introduced by time $t = \beta b$, and $\alpha b$ of those balls lie in bins of size at most $b$, we must choose $\beta$ such that $$b\beta - b\alpha \leq \int_{0}^{\beta b} (s_b(t))^d dt \leq \beta b(\alpha b)^{(d^b-1)\left(1 + \tfrac{1}{d-1}\right)}.$$ Dividing by $b\alpha$ and weakening the bound to simplify the expression gives the desired result. 

\section{End-to-end design and proofs}\label{sec:endtoend}

In this section, we describe how to use PSI-WCA to perform context-aware private contact tracing. Our implementation includes a set of isolated backend servers that will store and perform processing on the infected tokens as well as a client application that consists of a trusted and untrusted component. The untrusted app component will run outside the client's Trusted Execution Environment (TEE) while the the trusted app (TA) component will run inside the client's TEE and will guarantee authenticity of client data an queries.  In the rest of this section, unless explicitly stated otherwise, we use the term "client" to refer to the TA component of the client app. We assume that time and location information cannot be spoofed inside the TEE and that only the TA has access to data it stores in secure storage on the device.

\subsection{Bootstrapping}

\begin{figure}[hbt]
\centering
\includegraphics[scale=0.7]{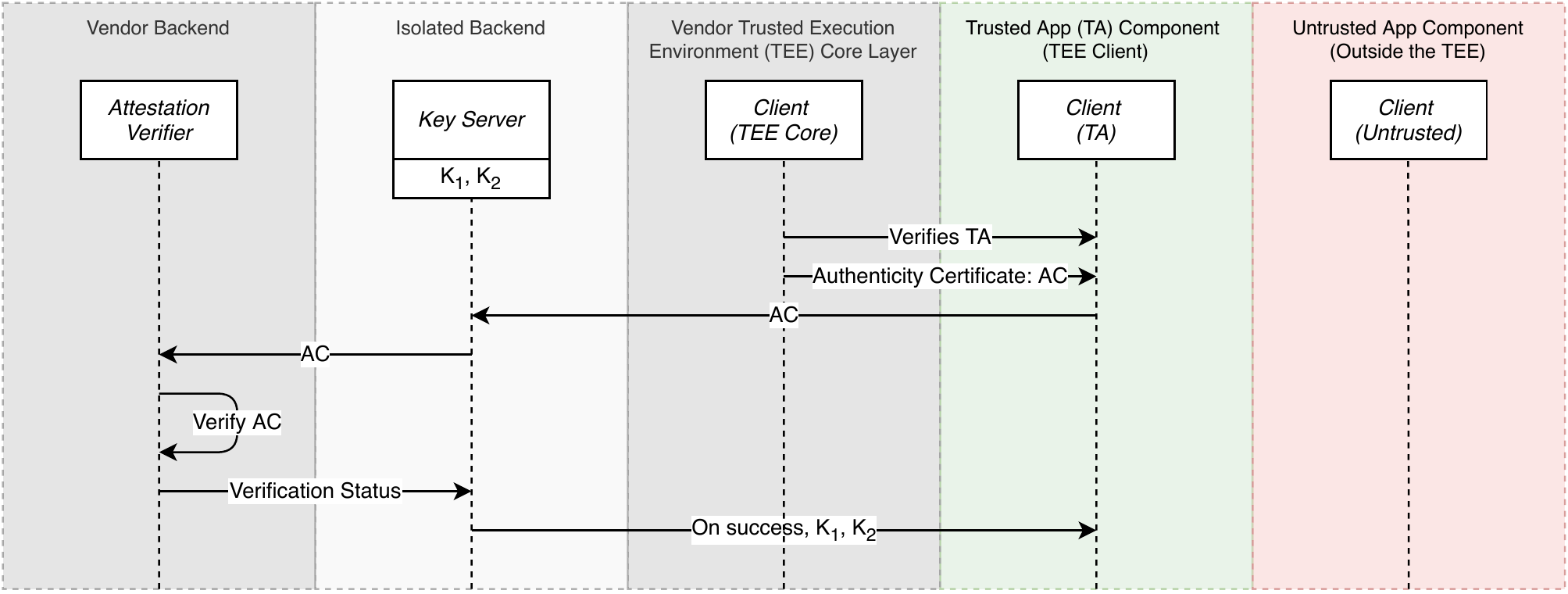}
\caption{Remote Attestation Overview}
\label{fig:remoteattestation}
\end{figure}

A session starts by the client by initiating a remote attestation request to prove the integrity of itself and its execution environment to the backend Key Server. In response, the client reecives two cryptographic keys $K_1$ and $K_2$ from the Key Server via the TEE vendor provided remote attestation process. A generic overview of this process is shown in Figure \ref{fig:remoteattestation}.

Here, key $K_1$ is shared only between the Key Sever and the client app. The purpose of $K_1$ is to prevent emulation of valid hashes outside the TA (if performed without knowing $K_1$, the hashes will not match with intended targets at the FSS servers, rendering any inferred information useless). Key $K_2$ is shared between the TA and the Submission Verification Server (VS). The purpose of $K_2$ is to guarantee authenticity of submissions to VS since the verification challenge passes through the untrusted app component. The detail of remote attestation is beyond the scope of this work. Implementation wise, the client app can utilize other vendor-specific remote attestation technologies that can guarantee runtime integrity of the client app (including its execution environment) without necessarily deploying a TA component inside a TEE. Key sharing from the Key Server to the client app is contingent upon successful remote attestation of the app's runtime integrity. Communication between protocol endpoints occurs over secure transport channels (e.g., over standard TLS with TLS public keys pinned at each endpoint). Transport security here is necessary to guarantee confidentiality and integrity of data in transient, orthogonal to the (internal) attestation and keying primitives used in the protocol itself which are needed to guarantee client non-tampering with its data and queries.

\subsection{Protocols}
\label{subsec:protocols}

In this subsection, we detail four protocols that are used to perform full end-to-end contact tracing. We first outline additional notation.

Let $\mathcal{T}$ be a secure token-generation algorithm, i.e. it generates tokens as nonces with high entropy, and for our purposes, $\mathcal{T}$ can simply be the uniform distribution on tokens of length $\kappa$.  Let $F(k,x)$ be a pseudorandom function. Let $H$ be a collision-resistant hash function that outputs a $\lambda$ bit ``true'' token, i.e. the tokens used in the PSI-WCA protocol (we use these two ``token'' terms interchangeably, though we typically refer to the latter ``true token'' notion). Let $(E,D)$ be a symmetric-key AEAD encryption algorithm. Let $\Pi_{\mathsf{PSI-WCA}}$ denote the protocol for (streaming) PSI-WCA.  We let an \emph{epoch} be a defined sliding window period of time.

\begin{figure}[hbt]
\centering
\includegraphics[scale=0.7]{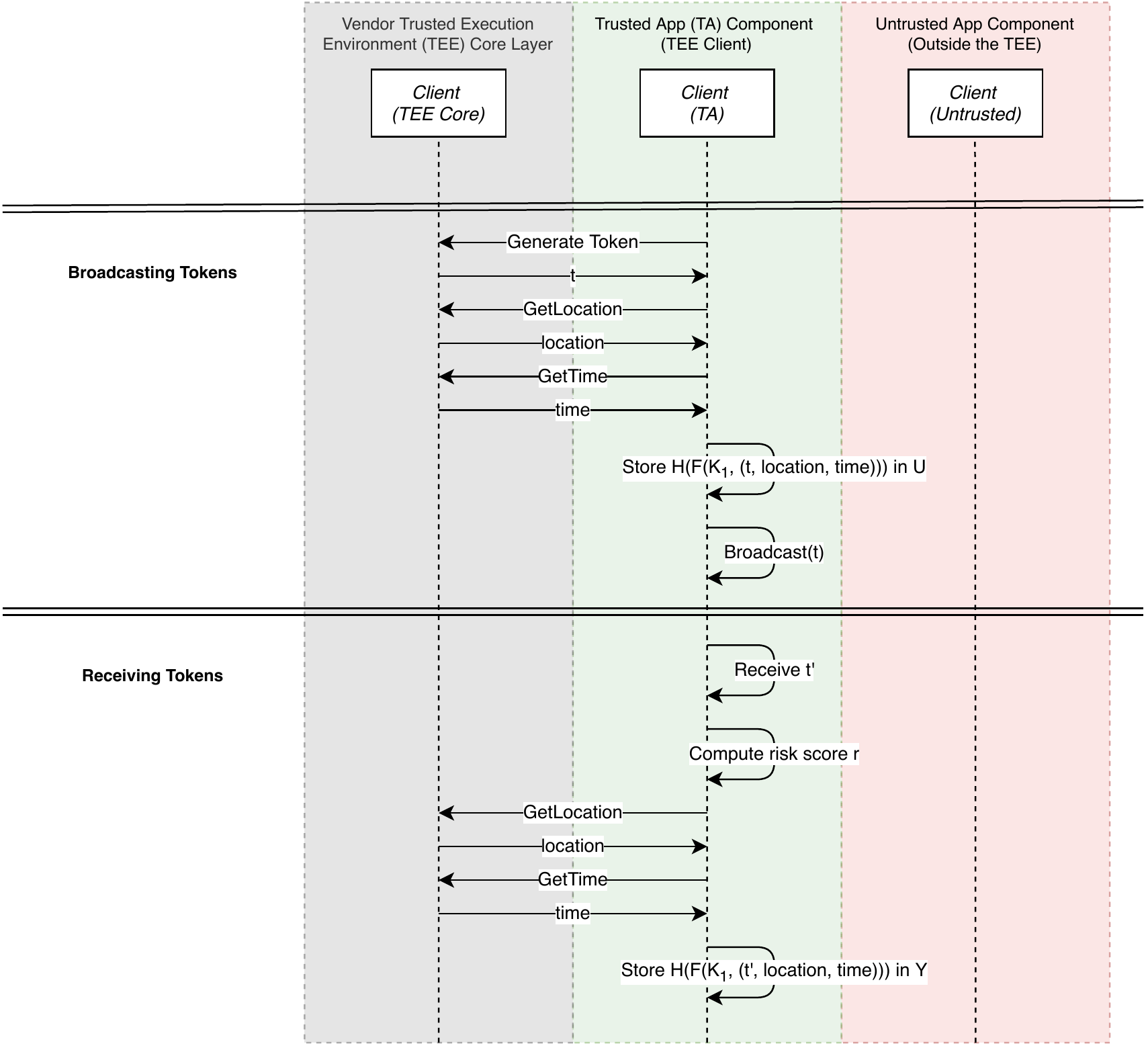}
\caption{Broadcasting and Receiving Tokens}
\label{fig:broadcastreceive}
\end{figure}

The process of token broadcast and receipt is outlined in Figure~\ref{fig:broadcastreceive}, where each client securely generates a token for broadcasting and calculates a hash of it based on the current location and timestamp. The full process is described in Algorithm~\ref{alg:broadcast} and Algorithm~\ref{alg:receive}. We also assume that time and location information cannot be spoofed inside the TEE.

\begin{algorithm}
\caption{\label{alg:broadcast} Protocol to broadcast a client token. }
\begin{algorithmic}[1]
\Procedure{Token Broadcast Protocol: $\Pi_{bcast}$}{}
\State The client generates a new token $t \gets \mathcal{T}$ at regular intervals. 
\State The client computes and stores the hash $u=H(F(K_1, (t, location, time)))$ in bucket $U$.
\State The client broadcasts $t$ to nearby devices.
\State (Upkeep) Tokens older than the epoch are discarded.
\EndProcedure
\end{algorithmic}
\end{algorithm}

\begin{algorithm}
\caption{\label{alg:receive} Protocol to receive a broadcasted token. }
\begin{algorithmic}[1]
\Procedure{Token Receipt Protocol: $\Pi_{receive}$}{}
\State The client receives the token $t'$ and computes a risk score $r$ that is associated with the received token $t'$.
\State The client computes and stores the hash $y=H(F(K_1, (t', location, time))$ in bucket $Y$.
\State (Upkeep) Tokens older than the epoch are discarded.
\EndProcedure
\end{algorithmic}
\end{algorithm}

\begin{figure}
\centering
\includegraphics[scale=0.7]{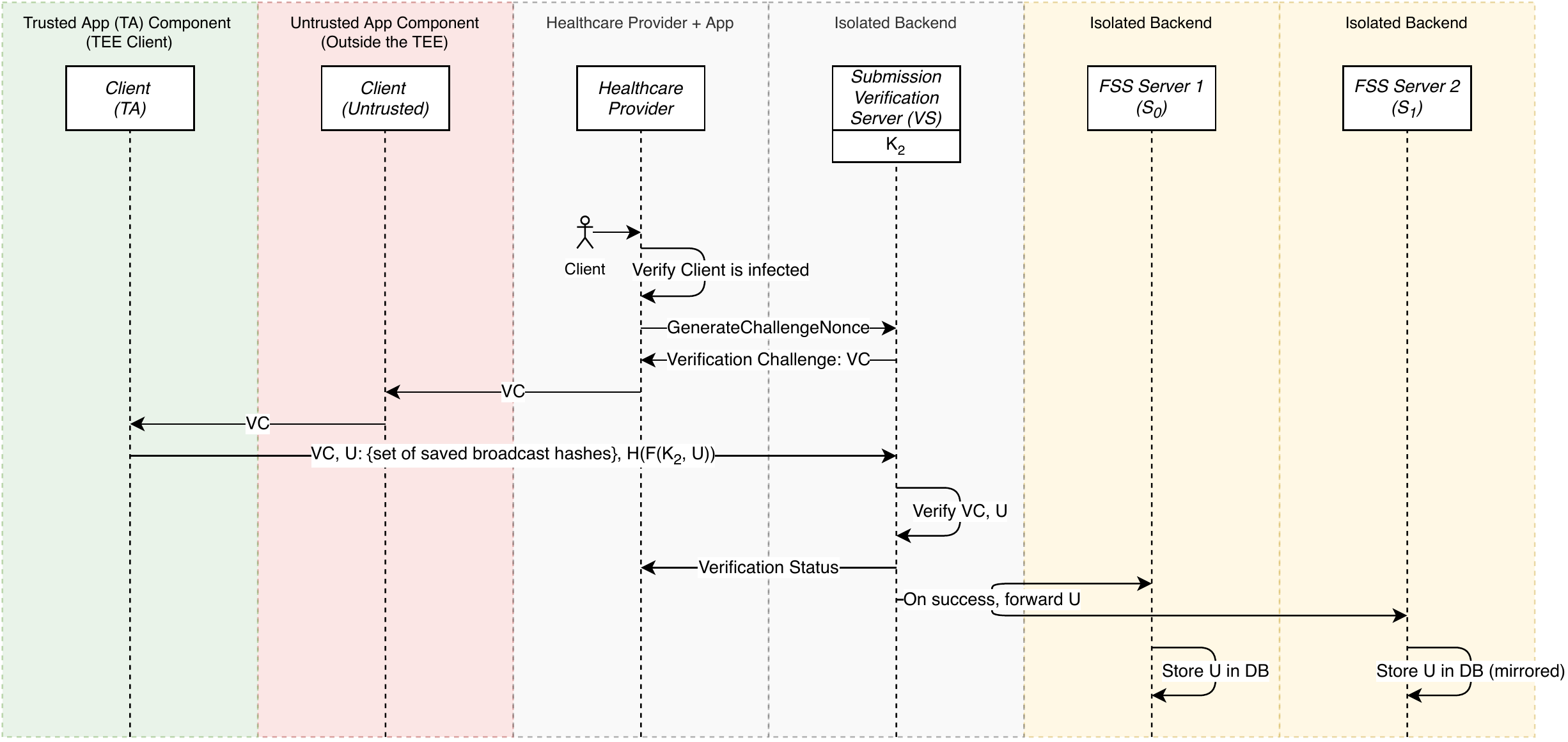}
\caption{Reporting Infection}
\label{fig:reportinfection}
\end{figure}

The process of reporting an infection is outlined in Figure~\ref{fig:reportinfection}. Whenever users get tested, they will visit a healthcare provider that will verify if they are infected. If that is the case, then the healthcare provider will provide a verification challenge that will be used to sign the tokens that will upload be uploaded to the back end servers. This step is necessary in order to assert that the uploaded tokens have not been tampered with. The full process is described in Algorithm~\ref{alg:upload}.

\begin{algorithm}
\caption{\label{alg:upload} Protocol to upload all user-generated tokens from the device of an infected user. }
\begin{algorithmic}[1]
\Procedure{Token Upload Protocol: $\Pi_{upload}$}{}
\State The user visits a Healthcare provider that verifies that the user is infected.
\State The Health care provider initiates a request to the Submission Verification Server and retrieves a Verification Challenge (VC). Note that the retrieved VC is entered into the client's untrusted app component and is forwarded from there to the trusted app component.
\State The client generates $u*=H(F(K_2, U))$ where $U$ is the set of stored hashes that was calculated in Algorithm~\ref{alg:broadcast} and $K_2$ is the key that is shared with the Submission Verification Server.
\State The Submission Verification Server verifies that $U$ indeed hashes to $u*$ and, upon success, forwards it to FSS Servers $S_0$ and $S_1$.

\State Servers $S_0$ and $S_1$ store $U$ into their internal database $X$.
\State (Upkeep) Tokens in $X$ that are older than the epoch are discarded.
\EndProcedure
\end{algorithmic}
\end{algorithm}

\begin{figure}
\centering
\includegraphics[scale=0.7]{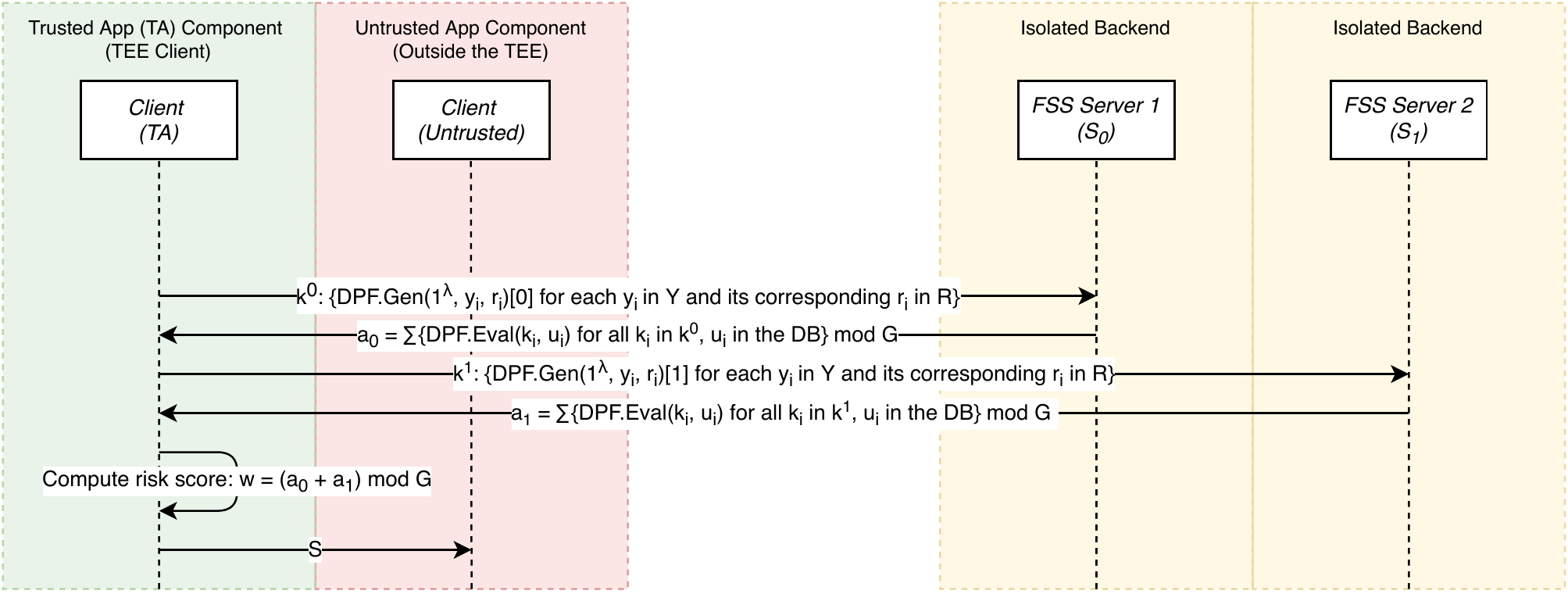}
\caption{Querying Risk Score}
\label{fig:queryriskscore}
\end{figure}

The process of calculating the risk score for each user is outlined in Figure~\ref{fig:queryriskscore}. In this step, the user will use each FSS server to calculate only part of the final risk score and will combine both replies to calculate the final risk score locally. The full process is described in Algorithm~\ref{alg:trace}. 

\begin{algorithm}
\caption{\label{alg:trace} Protocol to trace collected tokens against infected tokens to compute the risk score. }
\begin{algorithmic}[1]
\Procedure{Token Trace Protocol: $\Pi_{trace}$}{}
\State The client runs $\Pi_{\mathsf{PSI-WCA}}$ to generate the splits $k^0$ and $k^1$ by computing ${\mathsf{DPF.Gen}(1^\lambda, y_i, r_i)}$ for each $y_i \in Y$ and its corresponding risk score $r_i \in R$. 
\State The client sends $k_0$ to $S_0$ and $k_1$ to $S_1$.
\State $S_0$ runs $\Pi_{\mathsf{PSI-WCA}}$ and sends $a_0 = \sum_{k_i \in k^0, u_i \in X} \mathsf{DPF.Eval}(k_i, u_i) \mod G$ to the client.
\State $S_1$ runs $\Pi_{\mathsf{PSI-WCA}}$ and sends $a_1 = \sum_{k_i \in k^1, u_i \in X} \mathsf{DPF.Eval}(k_i, u_i) \mod G$ to the client.
\State The client computes the final risk score $w = (a_0 + a_1) \mod G$.

\EndProcedure
\end{algorithmic}
\end{algorithm}

\section{Security Analysis and Proofs}

\begin{theorem}
The set of protocols described in this section is secure in the TEE-enhanced malicious client, and two-party non-colluding semi-honest server model assuming the security of the PSI-WCA protocol, $(E,D)$, $F$, $K_1$, $K_2$, and the collision-resistance of $H$.
\end{theorem}

We consider the multiple forms of attacks that can be mounted by a client or a server in each protocol, and describe how our scheme mitigates them.  We also highlight some attacks we do not mitigate, and rule them out as trivially unavoidable, such as someone leaving their phone at home when traveling.

\subsection{Client-to-client Broadcast}

\smallskip\noindent\textsc{Omission Attacks.} A malicious client can always choose to not broadcast their token either by turning their phone off, putting it in a Faraday cage, or otherwise blocking the signal.  If this client was infected, this would generate false negatives during tracing.  We rule this out of scope.

\smallskip\noindent\textsc{Handoff Attacks.} A malicious client can always ask another person to carry their device for them.  If this client was infected, this would generate false positives during tracing.  We rule this out of scope.

\smallskip\noindent\textsc{Replaying Old Tokens.} A malicious client can re-broadcast their old tokens.  However, because the actual ``token'' used in PSI-WCA is the hash of the internal tokens along with the location and time, if the re-broadcast falls out of the same location or time period, then by the collision resistance of $H$, this will not collide with any real PSI-WCA token, and therefore is mitigated by the fact it will never intersect with anything.

\smallskip\noindent\textsc{Relaying Existing Tokens.} A malicious client can re-broadcast tokens it has received from other players.  Again, because the actual ``token'' used in PSI-WCA is the hash of the internal tokens along with the location and time, if the re-broadcast falls out of the same location or time period, then by the collision resistance of $H$, this will not collide with any real PSI-WCA token.  However, if it does quickly replay a token, it can effectively extend the ``infection strength'' of nearby devices.  First, this is not a very significant attack as it only amplifies nearby signals that clients would have most likely received from the legitimate source anyhow. Second, in order to counteract such tampering, we rely on the context-aware weights in PSI-WCA to apply meaningful heuristics to circumvent such amplification.

\smallskip\noindent\textsc{Fabricating Tokens.} Tokens can also be fabricated during broadcast by a malicious client.  However, this is mitigated by the collision resistance of $H$ as these will also not collide with any real PSI-WCA token and therefore will be ruled out.

\smallskip\noindent\textsc{Multi-device Attacks.} A malicious client or MITM can carry an enormous number of devices to amplify the signal of infected tokens.  This is mitigated in two ways.  First, the context-aware weights can heuristically determine that multiple tokens coming from roughly the exact same spot is suspicious and holds less weight. Second, the upload process of the devices is performed by a semi-honest healthcare provider: the malicious client would have trouble justifying to the healthcare agent why carrying an egregious number of devices is legitimate.

\smallskip\noindent\textsc{Non-human Handoff Attacks.} Placing the phone on an animal, stationary or mobile object is always an attack that can be mounted physically.  We mitigate this by using the context-aware weight to heuristically filter out non-human sources of token broadcasts.

\subsection{Client-to-client Receipt}

\smallskip\noindent\textsc{Omission Attacks.} A malicious client can always choose to not receive tokens broadcasted to them either by turning their phone off, putting it in a Faraday cage, or otherwise blocking the signal.  We rule this out of scope.

\smallskip\noindent\textsc{Handoff Attacks.} A malicious client can always ask another person to carry their device for them.  This is equivalent to the clients colluding, which reveal no additional information than the union of their knowledge.

\smallskip\noindent\textsc{Replay, Relay, or Fabrication of Tokens.} Any real or fake token inputted into the TEE will be location and timestamped, therefore if this does not match any real tuple, the collision resistance of $H$ will mitigate this attack.

\smallskip\noindent\textsc{Isolation and Multi-device Attacks.} A malicious client can interact with a single person and then perform contact tracing, which will reveal only the status of that person.  At a larger scale, a malicious client can carry multiple devices where throughout the day different subsets are turned on or off, and can learn the infected status of certain persons or groups of people via compressed sensing.  These attacks can be mounted against even an ideal functionality and cannot be prevented unless tracing intentionally adds errors to hide the result.  This tradeoff between privacy and utility is deemed to be in the scope of policy making, but our solution is compatible with the introduction of noise such as in differential privacy.

\smallskip\noindent\textsc{Non-human Handoff Attacks.} Placing the phone on an animal, stationary or mobile object is always an attack that can be mounted physically.  This allows the attacker to learn some potentially false information about the infection status of where the phone was due to there not being a human attached to it. We rule this out of scope.

\subsection{Client Upload}

\begin{lemma} The location and time of an honest client is hidden from the servers. \end{lemma}
\textit{Proof.}

Consider the distributions $\{U,u*\}$ and $\{U', u'*\}$ where $U'$ is obtained by hashing \emph{uniformly random} elements.  Clearly the latter distribution reveals nothing about the client's information.  We show $\{U,u*\} \stackrel{c}{\approx} \{U', u'*\}$ are computationally indistinguishable via a hybrid argument.  Replace $K_1$ and $K_2$ with random values, followed by replacing the $PRF$ output with random values.  These steps remain computationally indistinguishable under the security of the $K_1$, $K_2$, and $PRF$.
\qed

We now consider what a malicious client can perform to illicitly influence a server.

\smallskip\noindent\textsc{Omission Attacks.} A malicious client can always choose to not go to their healthcare provider to volunteer their tokens.  However, if they choose to volunteer their tokens, they cannot selectively omit tokens produced by the TEE because of the hash $u*$.

\smallskip\noindent\textsc{Handoff Attacks.} An infected malicious client can steal a device and pretend it is theirs or give their device to someone else who is infected to upload. We rule these out of scope.

\smallskip\noindent\textsc{Replay, Relay, or Fabrication of Tokens.} Any real or fake token that did not come directly from $U$ will be caught because the hash $u*$ will not match due to collision resistance. Note that in this case, collision resistance must hold even in the case of length-extension attacks.  Furthermore, if the symmetric-key encryption scheme works as a AEAD stream cipher, it will automatically authenticate the entire message without even having to worry about cut-and-paste attacks.

\smallskip\noindent\textsc{Security against eavesdroppers.} All messages between the TEE and the Server are encrypted, so even an eavesdropper on the phone itself cannot see anything.

\smallskip\noindent\textsc{Non-human Handoff Attacks.} Healthcare providers will not vouch for a non-human to upload, e.g. ``How did this dog get a phone?''

\subsection{Client Tracing}

\begin{lemma} The tokens of an honest client are hidden from the servers. \end{lemma}
\textit{Proof.}
This follows directly from the security of $\Pi_{\mathsf{PSI-WCA}}$. \qed

\begin{lemma} Even a malicious client cannot learn more than the weighted cardinality of the intersection between $Y$ and $X$. \end{lemma}
\textit{Proof.}
Because the last message from the client to the server is a hash of all previous messages, and it is encrypted under $sk$ (known only to the TEE), it serves as a binding ``committment'' of its previous messages.  If the last message of the malicious client is anything but a valid encryption of the hash, the servers will simply not respond and so the malicious client can be trivially simulated.  On the other hand, if it is a valid encryption of the hash, then it is infeasible for the malicious client to send anything but the valid (encrypted) set $Y$ that was produced out of the TEE.  This forces semi-honest behavior of the client in the PSI-WCA protocol, and therefore by the security of that protocol, the client only learns the output.
\qed

\smallskip\noindent\textsc{Omission Attacks.} Choosing not to run tracing is strictly less information provided to an attacker.  However, any token that was in $Y$ must have been there legitimately due to the security analysis of the client Broadcast/Receipt protocols.  The hash check at the end of the protocol ensures that any token $y\in Y$ (or rather, the FSS KeyGen ran on $y$) must have been included in that hash, so omission of it would result in the servers aborting.

\smallskip\noindent\textsc{Handoff Attacks.} Since the protocol is performed on the device, handing off the device does nothing.

\smallskip\noindent\textsc{Attempting to perform arbitrary queries via insertion.} Any token that is not in $Y$ cannot be inserted by a malicious attacker.  This is due to, again, the TEE providing a hash of all the messages it sent. Attempts to insert another token (or rather, FSS key of that token) would result in the hash check failing. 

\smallskip\noindent\textsc{Security against eavesdroppers.} All messages between the TEE and the Server are encrypted, so even an eavesdropper on the phone itself cannot see anything.

\smallskip\noindent\textsc{Non-human Handoff Attacks.} This attack is meaningless since the protocol is performed on the device, regardless of who is actually holding it.

\section{Comparison to Existing Schemes}

We summarize comparisons to other existing schemes.  
Of the most relevant schemes to highlight, we consider the Berkeley Epione proposal and the Apple and Google approach.  

We logically organize this discussion around the structure of token-based solutions. 
At a mile-high view, there are two types of stakeholders -- phones and servers.  
We identify three main workflows between these stakeholders: Phone-to-phone contact, Infected Phone Upload, Phone Query.  
Phone-to-phone contact is what happens when two phones come in contact with each other.  
Infected Phone Upload is the process which takes place when a person who is diagnosed to have the infection uploads data from their phone to the servers.  
Phone Query is what happens when a phone user wishes to query the servers to determine if they have come in contact with an infected user.

The manner in which tokens are generated we leave opaque to our system, and we treat them agnostic to their actual implementation.  
In order to allow for succinctness of revealing infected tokens to the server, it would be convenient to have them be generated pseudorandomly from a small seed, though this is not necessary.
We mention that there are advances in that area, and generating tokens correctly is a critical part of an overall solution.

\subsection{Apple and Google}

We summarize their solution, which primarily focuses on token generation and passing.  Let $t$ be Unix Epoch Time, and let $ENIN_t$ be a 32-bit little-endian unsigned integer representing the number of 10-minute intervals that have passed since January 1, 1970.  Let $TEKRollingPeriod$ denote how long a temporary key used to generate tokens is valid for, written as a multiple of 10 minutes (e.g. 144 is one day).  Then let $tek_i$ denote the 16-byte uniformly random key generated for valid window $i$ (e.g. for one day).

Let $H$ denote a secure HKDF (as defined by Krawczyk-Eronen in RFC5869) and define $RPIK_i=H(tek_i;\textsf{``EN-RPIK''})$, where the salt is omitted and the output is 16 bytes.

For day $i$ and 10-minute window $j$, define $RPI_{i,j}=AES128(RPIK_i;PaddedData_j)$ where $PaddedData_j = \textsf{``EN-RPI''||0x00 00 00 00 00 00||}ENIN_j$

Then $RPI_{i,j}$ is the token that is broadcast over Bluetooth. A metadata key can also be derived from $tek_i$ and encrypted metadata can also be sent over Bluetooth.

From the phone's perspective:

\begin{itemize}
\item \textbf{Phone-to-phone contact}: A token $RPI_{i,j}$ is passed from one phone to another.
\item \textbf{Infected Phone Upload}: From an infected phone, $tek_i$ is uploaded to a server for each day $i$ for the past 14 days.
\item \textbf{Phone Query}: From any phone, a list of all infected $\{tek_i\}$ is obtained from the server for each day $i$ for the past 14 days.  The phone then derives $RPIK_i$ from $tek_i$ then derives $RPI_{i,j}$ for each 10-minute interval $j$ for each of those infected keys.  It compares the output of all of these $RPI$ tokens to tokens it has seen over Bluetooth.
\end{itemize}

From the server's perspective:

\begin{itemize}
\item \textbf{Phone-to-phone contact}: Nothing is transmitted.
\item \textbf{Infected Phone Upload}: When a phone is discovered to be infected, $tek_i$ is uploaded to me for each day $i$ for the past 14 days.
\item \textbf{Phone Query}: I push out to every phone the list of all infected $\{tek_i\}$ for each day $i$ for the past 14 days.
\end{itemize}

This approach, while highly favorable to client-server communication, has privacy and security drawbacks, many of which have been pointed out by various researchers.
These range from linkability (if I have a good idea where I got a token from, I know whether or not it's infected) to malicious relay/replay attacks.
In essence, this solution is just to have each phone generate a random pseudonym every 10 minutes and locally beam it over Bluetooth, and infected pseudonyms in the past 14 days are made public to all phones.
This solution reveals more than just hit/miss: it reveals where the hits are!  To account for this, the Trieu et al. solution proposes a new PSI-CA solution that allows them to just get a count of the number of hits, and not where they are.
We detail this approach next.

\subsection{Epione (Berkeley)}

This solution takes the token generation and infected phone upload as a given starting point.  They consider the construction of an asymmetric-set-size PSI-CA as their novel contribution.  First, they consider the following Diffie-Hellman based solution to PSI-CA. Let the Server have input $X=\{x_1,\ldots,x_N\}$ and the Client have input $Y=\{y_1,\ldots,y_n\}$.  
$H$ be a random oracle that maps to some prime order $p$ group.  Then

\begin{enumerate}
\item Have the Server sample $\alpha$ and the Client sample $\beta$ uniformly from $\mathbb{F}_p$.  
\item The Client computes $m_i=H(y_i)^\beta$, sends it to the Server, and the Server computes $m_i'=m_i^\alpha$ and sends them back in \emph{randomly permuted order}.  
\item The Client then computes $v_i=(m_i')^{1/\alpha}$, now in unknown permuted order.  
\item The Server also computes and sends $u_i=H(x_i)^\beta$ randomly permuted to the client.  
\item The Client can now output the cardinality of $|\{v_i\}\cap\{u_i\}|$
\end{enumerate}

To reduce the communication when $n \ll N$, they provide an asymmetric solution using multi-query keyword PIR.  The observation is that Step 5 above can be replaced by the Client performing Keyword PIR with $v_i$ as the input.  Their paper then instantiates Keyword PIR using either 1-PIR or 2-PIR via Cuckoo Hashing and FSS.

In contrast, our solution only requires one round instead of two, and we use FSS directly to perform keyword search rather than use Keyword PIR.  Note that the Epione solution hides which tokens were hits and which were misses by having the server permute then blind them with an exponent. In our solution, the server uses the natural linearity of the FSS to sum up the counts \emph{before} they are sent to the Client.  This results in greatly reduced downstream communication.  Furthermore, our solution supports the ability for the client to supply weights to obtain a weighted cardinality.

\subsection{Table of Comparisons}

Tables~\ref{tab:cmp_p2p_upload} and \ref{tab:key_diff_phone_query} show the key differences between our proposed approach and existing solutions.

\begin{table}[hpt]
\begin{tabular}{|l|l|l|l|l|l|}
\hline
  & \multicolumn{2}{c|}{Phone-to-Phone Contact}   & \multicolumn{3}{c|}{Infected Phone Upload}  \\ \hline
  & Comp(P) & Comm $P\rightarrow S$ & Comp (P) & Comp (S) & Comm $P\rightarrow S$   \\ \hline
Apple \& Google & Gen Token & Token  & 0    & 0    & daily keys  \\ \hline
Epione & Gen Token & Token  & 0    & 0    & daily keys   \\ \hline
\textbf{Ours} & Gen Token & Token  & 0    & 0    & daily keys or raw tokens\\ \hline
\end{tabular}
\caption{Similarity of Phone-to-Phone Contact and Infected Phone Upload}
\label{tab:cmp_p2p_upload}
\end{table}

\begin{table}[hpt]
\begin{adjustwidth}{-1.8cm}{-1.8cm}
\begin{tabular}{|l|l|l|l|l|l|}
\hline
  & \multicolumn{5}{c|}{Phone Query}  \\ \hline
  & Comp(P)   & Comp(S) & Comm $P\rightarrow S$  & Comm $S\rightarrow P$    & Rounds \\ \hline
Apple \& Google & 0    & 0 & 0   & 14N keys  & 0.5    \\ \hline
Epione & n exp + & n exp +  & n group elements + & n group elements +  & 2  \\ 
 &  n KPIR(N).Gen & n KPIR(N).Eval &  n  KPIR queries &  n  KPIR responses &   \\ 
 &  $\approx n \textrm{ exp}+n\lambda \textrm{ AES}$ & $\approx n \textrm{ exp}+nN\log{N}\textrm{ AES}$& $\approx n \textrm{ gp.}+n\lambda \log{N} $ &  $\approx n \textrm{ gp.}+ 3 n\lambda \textrm{ bits}$&   \\ 
\hline
\textbf{Ours} (baseline) & n FSS.Gen   & n FSS.Eval on N tokens & n FSS keys & 1 group element  & 1  \\ 
     \S~\ref{ss:oneshot} & $\approx n\lambda \textrm{ AES}$ & $\approx nN\lambda \textrm{ AES}$  &  $\approx n\lambda \textrm{ |AES|}$ & (indep. of n)   &   \\ \hline
\textbf{Ours}  (queueing) & $c n$ FSS.Gen   & $b c$ FSS.Eval on N tokens & $\alpha^{-1} n$ FSS keys & 1 group element  & 1  \\ 
\S~\ref{ss:bucketing_design}, Table~\ref{tab:experiments} & $\approx c n \lambda \textrm{ AES}$ & $\approx b c N \lambda \textrm{ AES}$  &  $\approx \alpha^{-1} n\lambda \textrm{ |AES|}$ & (indep. of n)   &  \\ \hline
\end{tabular}
\caption{Key differences in Phone Query}
\label{tab:key_diff_phone_query}
\end{adjustwidth}
\end{table}

\section{Conclusion}\label{sec:conclusion}

In summary, we presented a new approach to PSI-Cardinality where we used 2-server FSS and extended it to ``streaming'' cardinality and Weighted Cardinality with applications to Contact Tracing.  We provided a description of an end-to-end protocol and analyzed its security against various forms of theoretical and practical attacks.

\bibliographystyle{alpha}

\newcommand{\etalchar}[1]{$^{#1}$}
\begin{thebibliography}{KKRT16}

\bibitem[ABKU94]{azar1994balanced}
Yossi Azar, Andrei~Z Broder, Anna~R Karlin, and Eli Upfal.
\newblock Balanced allocations.
\newblock In {\em Proceedings of the twenty-sixth annual ACM symposium on
  Theory of computing}, pages 593--602, 1994.

\bibitem[ACLS18]{AngelCLS18}
Sebastian Angel, Hao Chen, Kim Laine, and Srinath T.~V. Setty.
\newblock {PIR} with compressed queries and amortized query processing.
\newblock In {\em 2018 {IEEE} Symposium on Security and Privacy, {SP} 2018,
  Proceedings, 21-23 May 2018, San Francisco, California, {USA}}, pages
  962--979. {IEEE} Computer Society, 2018.

\bibitem[BGI15]{EC:BoyGilIsh15}
Elette Boyle, Niv Gilboa, and Yuval Ishai.
\newblock Function secret sharing.
\newblock In Elisabeth Oswald and Marc Fischlin, editors, {\em Advances in
  Cryptology -- {EUROCRYPT}~2015, Part~II}, volume 9057 of {\em Lecture Notes
  in Computer Science}, pages 337--367, Sofia, Bulgaria, April~26--30, 2015.
  Springer, Heidelberg, Germany.

\bibitem[BGI16]{CCS:BoyGilIsh16}
Elette Boyle, Niv Gilboa, and Yuval Ishai.
\newblock Function secret sharing: Improvements and extensions.
\newblock In Edgar~R. Weippl, Stefan Katzenbeisser, Christopher Kruegel,
  Andrew~C. Myers, and Shai Halevi, editors, {\em ACM CCS 2016: 23rd Conference
  on Computer and Communications Security}, pages 1292--1303, Vienna, Austria,
  October~24--28, 2016. {ACM} Press.

\bibitem[BW94]{bruneel1994analysis}
Herwig Bruneel and Ilse Wuyts.
\newblock Analysis of discrete-time multiserver queueing models with constant
  service times.
\newblock {\em Operations Research Letters}, 15(5):231--236, 1994.

\bibitem[CBM15]{CorriganGibbsB15}
Henry Corrigan{-}Gibbs, Dan Boneh, and David Mazi{\`{e}}res.
\newblock Riposte: An anonymous messaging system handling millions of users.
\newblock In {\em 2015 {IEEE} Symposium on Security and Privacy, {SP} 2015, San
  Jose, CA, USA, May 17-21, 2015}, pages 321--338. {IEEE} Computer Society,
  2015.

\bibitem[CGN98]{ChorGN98}
Benny Chor, Niv Gilboa, and Moni Naor.
\newblock Private information retrieval by keywords.
\newblock {\em {IACR} Cryptol. ePrint Arch.}, 1998:3, 1998.

\bibitem[CHLR18]{ChenHLR18}
Hao Chen, Zhicong Huang, Kim Laine, and Peter Rindal.
\newblock Labeled {PSI} from fully homomorphic encryption with malicious
  security.
\newblock In {\em {ACM} {CCS} 2018}, pages 1223--1237, 2018.

\bibitem[CLR17]{ChenLR17}
Hao Chen, Kim Laine, and Peter Rindal.
\newblock Fast private set intersection from homomorphic encryption.
\newblock In {\em {ACM} {CCS} 2017}, pages 1243--1255, 2017.

\bibitem[DFL{\etalchar{+}}20]{DORY}
Emma Dauterman, Eric Feng, Ellen Luo, Raluca~Ada Popa, and Ion Stoica.
\newblock Dory: An encrypted search system with distributed trust.
\newblock Cryptology ePrint Archive, Report 2020/1280, 2020.
\newblock \url{https://eprint.iacr.org/2020/1280}.

\bibitem[DPT20]{AC:DuoPhaTri20}
Thai Duong, Duong~Hieu Phan, and Ni~Trieu.
\newblock Catalic: Delegated psi cardinality with applications to contact
  tracing.
\newblock In {\em Asiacrypt 2020}, pages 1055--1072, 12 2020.

\bibitem[Eis08]{eisenberg2008expectation}
Bennett Eisenberg.
\newblock On the expectation of the maximum of iid geometric random variables.
\newblock {\em Statistics \& Probability Letters}, 78(2):135--143, 2008.

\bibitem[FNP04]{FreedmanNP04}
Michael~J. Freedman, Kobbi Nissim, and Benny Pinkas.
\newblock Efficient private matching and set intersection.
\newblock In {\em {EUROCRYPT} 2004}, pages 1--19, 2004.

\bibitem[GI14]{EC:GilIsh14}
Niv Gilboa and Yuval Ishai.
\newblock Distributed point functions and their applications.
\newblock In Phong~Q. Nguyen and Elisabeth Oswald, editors, {\em Advances in
  Cryptology -- {EUROCRYPT}~2014}, volume 8441 of {\em Lecture Notes in
  Computer Science}, pages 640--658, Copenhagen, Denmark, May~11--15, 2014.
  Springer, Heidelberg, Germany.

\bibitem[GN19]{GhoshN19}
Satrajit Ghosh and Tobias Nilges.
\newblock An algebraic approach to maliciously secure private set intersection.
\newblock In {\em {EUROCRYPT} 2019, Part {III}}, pages 154--185, 2019.

\bibitem[IKN{\etalchar{+}}17]{IonKNPSSSY17}
Mihaela Ion, Ben Kreuter, Erhan Nergiz, Sarvar Patel, Shobhit Saxena, Karn
  Seth, David Shanahan, and Moti Yung.
\newblock Private intersection-sum protocol with applications to attributing
  aggregate ad conversions.
\newblock {\em {IACR} Cryptol. ePrint Arch.}, 2017:738, 2017.

\bibitem[IKOS04]{IshaiKOS04}
Yuval Ishai, Eyal Kushilevitz, Rafail Ostrovsky, and Amit Sahai.
\newblock Batch codes and their applications.
\newblock In L{\'{a}}szl{\'{o}} Babai, editor, {\em Proceedings of the 36th
  Annual {ACM} Symposium on Theory of Computing, Chicago, IL, USA, June 13-16,
  2004}, pages 262--271. {ACM}, 2004.

\bibitem[JVL05]{janssen2005analytic}
Augustus~JEM Janssen and JSH Van~Leeuwaarden.
\newblock Analytic computation schemes for the discrete-time bulk service
  queue.
\newblock {\em Queueing Systems}, 50(2-3):141--163, 2005.

\bibitem[KKRT16]{KolesnikovKRT16}
Vladimir Kolesnikov, Ranjit Kumaresan, Mike Rosulek, and Ni~Trieu.
\newblock Efficient batched oblivious {PRF} with applications to private set
  intersection.
\newblock {\em {IACR} Cryptol. ePrint Arch.}, 2016:799, 2016.

\bibitem[Lit61]{little1961proof}
John~DC Little.
\newblock A proof for the queuing formula: L= $\lambda$ w.
\newblock {\em Operations research}, 9(3):383--387, 1961.

\bibitem[Mea86]{Meadows86}
Catherine~A. Meadows.
\newblock A more efficient cryptographic matchmaking protocol for use in the
  absence of a continuously available third party.
\newblock In {\em Proceedings of the 1986 {IEEE} Symposium on Security and
  Privacy}, pages 134--137, 1986.

\bibitem[Mit99]{mitzenmacher1999studying}
Michael Mitzenmacher.
\newblock Studying balanced allocations with differential equations.
\newblock {\em Combinatorics, Probability and Computing}, 8(5):473--482, 1999.

\bibitem[OS07]{OstrovskyS07}
Rafail Ostrovsky and William~E. Skeith.
\newblock Private searching on streaming data.
\newblock {\em J. Cryptology}, 20(4):397--430, 2007.

\bibitem[PRTY19]{PinkasRTY19}
Benny Pinkas, Mike Rosulek, Ni~Trieu, and Avishay Yanai.
\newblock Spot-light: Lightweight private set intersection from sparse {OT}
  extension.
\newblock In {\em {CRYPTO} 2019, Part {III}}, pages 401--431, 2019.

\bibitem[SGRR19]{CCS:SGRR19}
Phillipp Schoppmann, Adri{\`a} Gasc{\'o}n, Leonie Reichert, and Mariana
  Raykova.
\newblock Distributed vector-{OLE}: Improved constructions and implementation.
\newblock In {\em ACM CCS 2019: 26th Conference on Computer and Communications
  Security}, pages 1055--1072. {ACM} Press, 2019.

\bibitem[TSS{\etalchar{+}}20]{abs-2004-13293}
Ni~Trieu, Kareem Shehata, Prateek Saxena, Reza Shokri, and Dawn Song.
\newblock Epione: Lightweight contact tracing with strong privacy.
\newblock {\em CoRR}, abs/2004.13293, 2020.

\end{thebibliography}
\newcommand{\etalchar}[1]{$^{#1}$}

\end{document}